\documentclass[twocolumn,prb, superscriptaddress]{revtex4}
\usepackage{amsmath,graphicx,amsfonts,xcolor,float}

\usepackage{soul}

\begin{document}
\title{Correlation function diagnostics for type-I fracton phases}
\author{Trithep Devakul}
\affiliation{Department of Physics, Princeton University, Princeton NJ 08540, USA}
\author{S. A. Parameswaran}
\altaffiliation{On leave from: Department of Physics and Astronomy, University of California Irvine, Irvine CA 92617, USA.}
\affiliation{The Rudolf Peierls Centre for Theoretical Physics, University of Oxford, Oxford OX1 3NP, UK}
\author{S. L. Sondhi}
\affiliation{Department of Physics, Princeton University, Princeton NJ 08540, USA}

\begin{abstract}

Fracton phases are recent entrants to the roster of topological phases in three dimensions. They
are characterized by subextensively divergent topological degeneracy and excitations that are
constrained to move along lower dimensional subspaces, including the eponymous fractons that are
immobile in isolation. We develop correlation function diagnostics to characterize Type I fracton
phases which build on their exhibiting {\it partial deconfinement}.  
These are inspired by similar diagnostics from standard gauge theories and utilize a generalized
gauging procedure that links fracton phases to classical Ising models with subsystem symmetries. 
En route, we explicitly construct the spacetime partition function for the plaquette Ising model which, under such gauging,
maps into the X-cube fracton topological phase.
We numerically verify our results for this model via Monte Carlo calculations.

\end{abstract}
\maketitle

\emph{Introduction.}---Recent studies~\cite{haah,chamon,bravyi,yoshida,vijay2,williamson,vijay}
of exactly-solvable stabilizer codes in three dimensions have identified a new class of topologically ordered states that exhibit subextensive topological degeneracy on closed manifolds. Unlike the emergent gauge theories of topological order these ``fracton'' models lack a point-like excitation free to propagate in 3D. Owing to this, they exhibit translationally-invariant glassy dynamics even at nonzero energy density~\cite{kimhaah, nandkishore}.
Instead of fully deconfined point particles, their excitation spectrum generically includes immobile ``fractons'',
as well as a hierarchy of other excitations free to move along lower-dimensional subspaces. Depending on whether fractons may be created at the corners of two-dimensional membranes, or only upon the application of fractal operators, fracton models may be further divided into `Type I' or `Type II' fracton phases, in turn related to distinct subsystem symmetries of the classical spin models related to them via a generalized gauging procedure~\cite{williamson,vijay}.
%Finally we note that the resonating plaquette phase discussed in Ref.~\cite{pankov,cenke} appears to be a strongly coupled fracton phase much
%as the resonating valence bond phase is the deconfined phase at strong coupling in an odd Ising gauge theory~\cite{moessner}.
Finally we note that resonating plaquette phases as discussed in Ref.~\onlinecite{pankov,cenke,fccrsvp} have the potential to describe fracton phases.

 Despite rapid progress~\cite{pretko0, pretko1, pretko,pretko2, pretko3, layered, layered2, kim, kim2, fractopartons, fractonchains, nandkishore2, petrova} in advancing the theory of these novel 3D topological phases, there is a paucity of sharp characterizations of fracton deconfinement away from the stabilizer limit, e.g. when fractons acquire dynamics or are at finite density. One possible diagnostic is to extract topological contributions to the entanglement entropy~\cite{luEE, FractonEE1,FractonEE2}, but this requires an exact computation of ground states, typically challenging in 3D, and does not immediately generalize to $T>0$. 
For topological orders described by standard lattice gauge theories,
 a trio of loop observables suitably oriented in Euclidean space-time serves this role, and furthermore may be directly computed from, e.g. quantum Monte Carlo simulations. 
 Can such diagnostics be adapted to study these new states in the presence of dynamical fractonic matter?

Here, we answer this in the affirmative for the so-called X-cube model, and argue that our results may be generalized to all  Type-I fracton phases of which it is the paradigmatic example. We do so by formulating a  generalized ``plaquette gauge theory'' (PGT)  for  the plaquette Ising model, a classical spin model with spin-flip symmetries along planar subsystems. The PGT (and its dual, which we will introduce) describes a perturbed X-cube model.
Although  quasiparticle excitations of these models are always constrained to  lower-dimensional subspaces and are hence not truly deconfined, they are in a sense \emph{partially} deconfined within these subspaces. 
We show that the standard technology for diagnosing the deconfined and confined phases~\cite{gregor,chandran}, reviewed next,  can indeed be generalized in a straightforward manner to detect this partial deconfinement that can be viewed as a defining property of fractonic matter.  

\textit{Ising Gauge Theory.}---
To orient our discussion, we first review the gauging procedure that leads to the Ising gauge theory (IGT), and discuss its deconfinement diagnostics~\cite{gregor}. We begin with the classical Ising Hamiltonian on the square lattice, with matter degrees of freedom $\tau^z_s$ on the site $s$, and nearest-neighbor $J\tau^z\tau^z$ interactions (we will often suppress the site subscript when the meaning is obvious).
This model has a global $\mathbb{Z}_2$ symmetry, which is a flip of all $\tau^z$, that can be `gauged' by introducing an Ising spin $\sigma^z_l$ on each link $l$, and modifying the interaction term accordingly: $J\tau^z\tau^z\rightarrow J\sigma^z\tau^z\tau^z$. 
This expands the global Ising symmetry to a local $\mathbb{Z}_2$ gauge symmetry $G_s$ on each site, obtained by considering a simultaneous flip of $\tau^z_s$ and each $\sigma^z$ coupled to it by an interaction term --- i.e., those on the 4 links surrounding site $s$.
The IGT is obtained by restricting to the subspace where $G_s=+1$ for all $s$.
Finally, we give quantum dynamics to  both gauge and matter degrees of freedom by adding terms $\Gamma\sigma^x$ and $\Gamma_M\tau^x$ to our Hamiltonian.
To complete our construction of the IGT Hamiltonian, we add a gauge  `potential energy' by identifying the  simplest gauge-invariant pure-$\sigma^z$ term that commutes with $\tau^x$, here a product of $\sigma^z$ around a plaquette $p$, with coupling strength $K$, yielding
\begin{eqnarray}
\mathcal{H}_{\text{IGT}} &=& -K\sum_{p}\prod_{l\in\partial p} \sigma^z_l - \Gamma_M\sum_{s}\tau^x_s \\
&& - J\sum_{l}\sigma^z_l\prod_{s\in\partial l}\tau^z_s- \Gamma\sum_{l}\sigma^x_l \nonumber
\end{eqnarray}
subject to the constraint
$G_s = \tau^x_s\prod_{l\in\partial s}\sigma^x_l = 1$,
where $s, l, p$ denote links, sites, and plaquettes, and we denote by $\partial s, \partial l, \partial p$ the objects touching them (in this case the 4 links surrounding a site, the 2 sites straddling a link, and the 4 links encircling a plaquette).

Precisely at $J=\Gamma=0$, this model reduces to Kitaev's Toric code~\cite{toriccode} (this can be seen by enforcing the constraint to replace $\tau^x_s$ by $\prod_{l\in\partial s}\sigma^x_l$).  
Introducing nonzero $J$ or $\Gamma$ can then be thought of as perturbations from the Toric code point.
Turning $\Gamma$ too high will drive the gauge theory into a trivial confined phase, and turning $J$ too high will result in a Higgs transition into a symmetry broken phase where $\langle \tau^z\rangle$ obtains an expectation value.
These two limits are smoothly connected~\cite{fradkinshenker}, thus we will refer to both as the confined limits, and small perturbations of the Toric code point as the deconfined limit (characterized by $\mathbb{Z}_2$ topological order).

Let us now consider moving along the ``pure gauge theory'' axis, $\Gamma>0, J=0$, along which the matter is static, $\tau^x_s=1$ and therefore can be ignored. 
Here, the spatial Wilson loop, $W=\prod_{l\in C}\sigma_l^z$, where $C$ is a closed loop (taken for simplicity to be an $L\times L$ square), serves as a diagnostic that can distinguish the confined and deconfined phases.  
At the Toric code point ${\Gamma}=0$, we have $\langle W\rangle=1$.
Small perturbations in ${\Gamma}$ create local fluctuations of pairs of ``visons'', plaquettes on which $\prod_{l\in\partial p}\sigma_l^z=-1$ (the magnetic flux excitations of the theory).
As the Wilson loop measures the average parity of visons contained within it, these fluctuations will cause the  expectation value to decay proportionally to the perimeter of the loop, following a perimeter law: $\log\langle W\rangle \sim -L$ for large $L$.
In the confined phase at large ${\Gamma}$, the visons are condensed and so here $\log\langle W \rangle \sim - L^2$ follows an \emph{area} law for large $L$. 
However, as soon as we add dynamical matter $J>0$, the Wilson loop follows a perimeter law \emph{everywhere}.
To see this, notice that in perturbation theory in ${J}$ about the $J=0$ ground state $|\psi_0\rangle$, a term matching the Wilson loop operator appears at $O(J^L)$: $|\psi\rangle=|\psi_0\rangle+\alpha e^{-\beta L} W |\psi_0\rangle + \dots$ for some numbers $\alpha\sim O(1)$ and $\beta\sim -\ln J$, so that there is at least a perimeter law component to $\langle W \rangle$ which dominates as $L\rightarrow\infty$.
Thus, the Wilson loop fails as a deconfinement diagnostic as soon as $J>0$.

Now, consider moving along the ``pure matter theory'' axis, with $J>0, \Gamma=0$.
Here, the gauge field exhibits no fluctuations, and it is convenient to work with $\sigma^z=1$, and project onto the gauge invariant subspace if needed. 
In this subspace, the Hamiltonian is simply the original Ising model, in a transverse field.  
Beyond a critical ${J}$, there is a transition to an ordered phase where $\langle \tau^z\rangle$ gains an expectation value.
However, $\tau^z$ alone does not correspond to a gauge invariant operator; only pairs of $\tau^z$ do.  
This transition can therefore be diagnosed by an \emph{open} Wilson loop $\tau^z_{s}\tau^z_{s^\prime}\prod_{l\in C_{s s^\prime}}\sigma^z_l$ where $C_{s s^\prime}$ is a path connecting sites $s$ and $s^\prime$, which in this subspace is simply the spin-spin correlation function $\langle \tau^z_s \tau^z_{s^\prime}\rangle$.
As one takes $|s-s^\prime|\rightarrow\infty$, this either goes to zero in the deconfined (paramagnetic) phase, or approaches a constant in the confined (Higgs ferromagnetic) phase.
This can also be understood without referring to the matter theory as the vanishing of a line-tension in the Euclidean action~\cite{gregor}.
Now consider adding in a small ${\Gamma}$ perturbatively: $\sigma^x$ anticommutes with the $\sigma^z$ chain, and so $\langle \tau^z_s\sigma^z\dots\sigma^z\tau^z_{s^\prime}\rangle$ decays to zero exponentially with $|s-s^\prime|$ in both phases.
We therefore again are in a situation where a diagnostic that works exactly along this axis fails as soon as ${\Gamma}>0$.

How then can we distinguish the confined from the deconfined phase away from these special axes?  
The answer is to measure an appropriate line tension, using wisdom gained from the Euclidean path integral representation which maps the problem on to an isotropic 3D statistical mechanical problem of edges and surfaces~\cite{gregor,huseleibler}.
This can be linked to the expectation value of a ``horseshoe operator'', viz.  an $L\times L$ Wilson loop cut in half (with $\tau^z$ inserted at the ends for gauge invariance), 
$W_{1/2} = \tau^z_s \tau^z_{s^\prime} \prod_{l\in C_{1/2}}\sigma^z_l$, 
where $C_{1/2}$ defines the half-Wilson loop of dimension $L/2\times L$, terminating at sites $s$ and $s^\prime$.
The ratio of expectation values as $L\rightarrow\infty$, 
\begin{eqnarray}
    \mathcal{R}(L) = \frac{\langle W_{1/2} \rangle}{\sqrt{\langle W \rangle}} \xrightarrow[]{L\rightarrow\infty}
    \begin{cases} 0 & \text{deconfined} \\
     \text{const.} & \text{confined}
    \end{cases}
    \label{eq:ratio}
\end{eqnarray}
 can then be understood as measuring the ``cost'' of opening the Wilson loop.
In the deconfined phase, opening a Wilson loop will cause the expectation value to decay exponentially with the size of the gap.
In the confined phase, the expectation value of the Wilson loop follows a perimeter law regardless of whether it is opened or closed, thus the scaling with $L$ is exactly cancelled out by dividing by the square root of the full Wilson loop.

Since the Euclideanized IGT is space-time symmetric, by choosing distinct orientations and `cuts' of the loop, we can identify three different diagnostics. Besides (1) the `spatial loop' discussed above, the two possible cuts for the orientation extending along the time direction also have elegant physical interpretations~\cite{gregor}: either (2)~as the Fredenhagen-Marcu diagnostic~\cite{fredenhagenmarcu1,fredenhagenmarcu2}, measuring the overlap between the ground state and the normalized two-spinon state; or (3) as a measure of delocalized spinon (electric-charge)  excitations.
By the self-duality of the IGT this  exercise could have been done in the dual model, which defines a different Wilson loop object and exactly interchanges the role of the gauge (${\Gamma}$, ${K}$) and matter (${J}$, ${\Gamma}_M$) sectors~\cite{kogut}.

\textit{Euclidean Path Integral and Wilson Loops for Plaquette Gauge Theory }.---
We will now proceed with our analysis of the ``plaquette gauge theory'' (PGT), which arises from applying the generalized gauging procedure to the classical plaquette Ising model~\cite{vijay,williamson,johnston1} and produces X-cube fracton topological order in its deconfined phase, by analogy with the IGT of the preceding section.
The main deviation from the standard gauging procedure is that we place $\sigma$  at the center of each interaction in the Hamiltonian (the plaquettes in this model), rather than always on the links (these are the ``nexus'' spins of Ref.~\onlinecite{vijay}).

The classical (3D) plaquette Ising model (CPIM) is described by $\mathcal{H}_\text{CPIM} = -J\sum_p\prod_{s\in\partial p}\tau_s^z$, where the sum is over plaquettes and the product is over the four sites at the corner of plaquette $p$.
Applying the gauging procedure, we arrive at the PGT Hamiltonian,
\begin{eqnarray}
\mathcal{H}_{\text{PGT}} &=& -K\sum_{c,i}\prod_{p\in b_i(c)} \sigma^z_p - \Gamma_M\sum_{s}\tau^x_s \label{eq:pgt} \\
&& - J\sum_{p}\sigma^z_p\prod_{s\in\partial p}\tau^z_s- \Gamma\sum_{p}\sigma^x_p \nonumber
\end{eqnarray}
where now the $\sigma$s  live at the center of  plaquettes $p$, $c$ denotes a cube, and $b_i(c)$ for $i=1,2,3$ correspond to the three distinct combinations of four plaquettes that wrap around the cube $c$ (sometimes aptly called ``matchboxes'').
We further have a constraint defined on each site $s$,
$G_s = \tau^x_s\prod_{p\in\partial s}\sigma^x_p = 1$,
where the product is over the 12 plaquettes touching $s$.
Note that this model, for small $J$ and $\Gamma$, is just a perturbed X-cube model (which is usually defined on the \emph{dual} lattice where our plaquettes become links)
and that the topological order is stable to small perturbations~\cite{stability}.
The deconfined phase of this model hosts two types of excitations: the ``electric'' ($\tau^x=-1$) excitations are fractons, while the ``magnetic'' excitations are one-dimensionally mobile quasiparticles, which we will refer to as lineons (short for ``line vison'').

In standard gauge theory, one is often only concerned about the deconfinement of the electric charge excitations.
The X-cube model (unlike the Toric code) does not possess an electro-magnetic ($\sigma^z\leftrightarrow\sigma^x$) self-duality, so for completeness we also consider the ``electromagnetic'' \emph{dual} to the PGT.
This dual model arises naturally from the same generalized gauging procedure on the \emph{classical} dual of the CPIM, which can be written as an anisotropically coupled Ashkin-Teller model~\cite{johnston2,ashkinteller}.
Note that the duality discussed here maps between two full gauge-matter theories; 
the ``F-S duality'' between a pure matter theory and pure fracton gauge theory~\cite{vijay} is a limiting case.
We construct deconfinement diagnostics for the electric charge in both the PGT and its dual, thus providing diagnostics for both fracton and lineon excitations.

For a full space-time discussion of Wilson loop analogues, we construct a discrete-time Euclidean path integral for the PGT Hamiltonian Eq.~\eqref{eq:pgt} via the usual Suzuki-Trotter decomposition.  
The gauge constraint is enforced by the introduction of auxiliary spin-$1/2$ degrees of freedom along the time-links of the 4D hypercubic lattice~\cite{kogut,moessner}, that we will denote $\lambda$ (in the IGT one has a space-time symmetric structure so these spins can be thought of as $\sigma$ spins along the time-links, but this is not the case here). After a straightforward calculation (for details, see~\cite{supmat}), we find $Z_{\text{PGT}}=\text{Tr}_{\left\{\tau,\sigma,\lambda\right\}} e^{-\mathcal{S}_{\text{PGT}}}$, with the Euclidean action
\begin{eqnarray}
    \mathcal{S}_{\text{PGT}}  &=& -\tilde{K}\sum_{t,c,i}\prod_{p\in b_i(c)} \sigma_p^{(t)} - \tilde{\Gamma}_M\sum_{t,s}\tau_s^{(t)}\lambda_s^{(t)}\tau_s^{(t+1)}
    \label{eq:pgtaction}
    \\
    && - \tilde{J}\sum_{t,p}\sigma^{(t)}_p\prod_{s\in\partial p}\tau_s^{(t)} - \tilde{\Gamma}\sum_{t,p}\sigma_p^{(t)}\sigma_p^{(t+1)}\prod_{s\in\partial p}\lambda_s^{(t)} \nonumber
\end{eqnarray}
where the integer $t$ labels the Euclidean time slice (which extends to infinity for zero temperature), $\tau_s^{(t)}$ ($\sigma_l^{(t)}$) is now a classical Ising variable associated with sites (links) in the time slice $t$, and $\lambda_s^{(t)}$ is similarly associated with the link connecting site $s$ between time slices $t$ and $t+1$.
The couplings in $  \mathcal{S}_{\text{PGT}} $ are related to those in $\mathcal{H}_{\text{PGT}}$ and the Trotter time step $\epsilon$ via
$\tilde{K} = \epsilon K$, $\tilde{J} = \epsilon J$, and $\tilde{\Gamma}_{(M)} = -\frac{1}{2}\log \tanh\epsilon\Gamma_{(M)}$.
This can be viewed as a statistical mechanical model of edges, surfaces, and volumes in 4D, but with a  more subtle set of rules for how to build allowed objects from these.

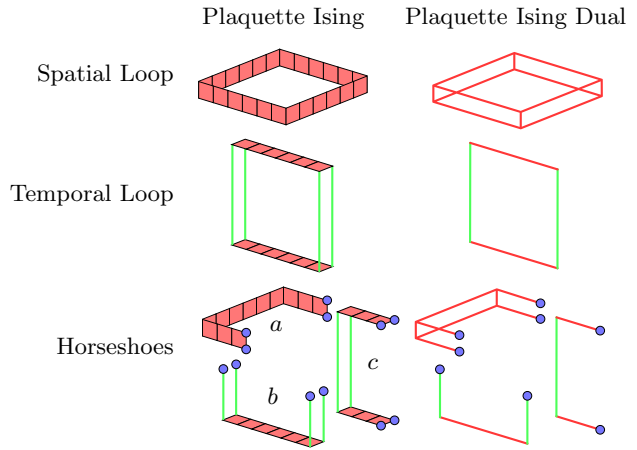
\begin{figure}[t]
    \centering
    \setlength{\unitlength}{267.84768066bp}%
    \begin{picture}(1,0.67648742)%
        \put(0,0){\includegraphics[width=\unitlength,page=1]{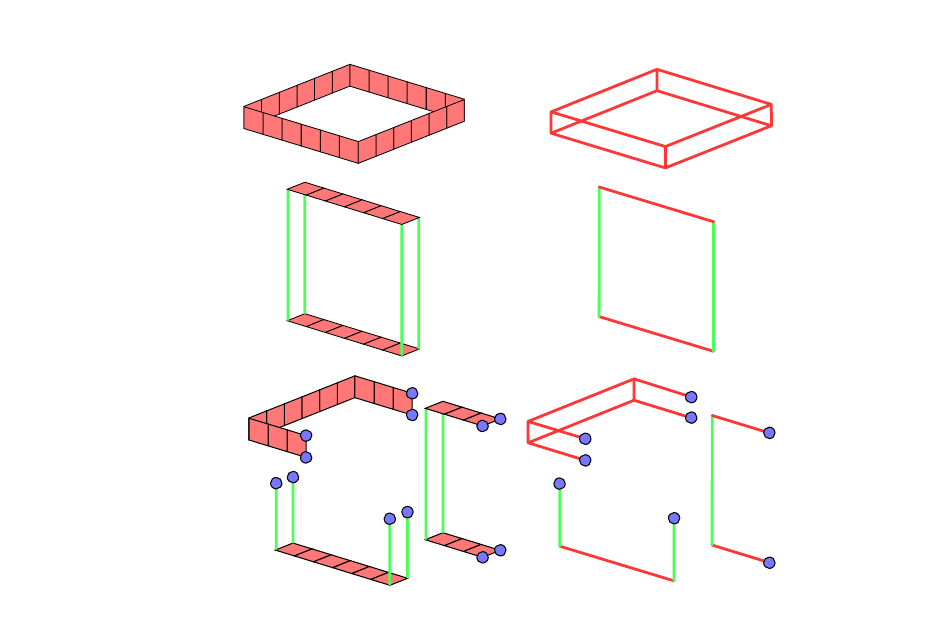}}%
        %\put(0.260,0.050){\includegraphics[width=0.55\unitlength,page=1]{loops.eps}}%
        \put(0.50007538,0.15637237){\color[rgb]{0,0,0}\makebox(0,0)[lb]{\smash{$c$}}}%
        \put(0.35904369,0.10773241){\color[rgb]{0,0,0}\makebox(0,0)[lb]{\smash{$b$}}}%
        \put(0.36223655,0.20986675){\color[rgb]{0,0,0}\makebox(0,0)[lb]{\smash{$a$}}}%
        \put(0.26739641,0.64244872){\color[rgb]{0,0,0}\makebox(0,0)[lb]{\smash{Plaquette Ising}}}%
        \put(0.55283733,0.64244872){\color[rgb]{0,0,0}\makebox(0,0)[lb]{\smash{Plaquette Ising Dual}}}%
        \put(0.22745335,0.56251318){\color[rgb]{0,0,0}\makebox(0,0)[rb]{\smash{Spatial Loop}}}%
        \put(0.22745335,0.39379989){\color[rgb]{0,0,0}\makebox(0,0)[rb]{\smash{Temporal Loop}}}%
        \put(0.22745335,0.1780056){\color[rgb]{0,0,0}\makebox(0,0)[rb]{\smash{Horseshoes}}}%
    \end{picture}%
    \caption{The Euclidean time representation of the Wilson loop and horseshoe generalizations for the PGT and its dual, which realize the X-cube topological phase.
        Blue circles represents $\tau$ (which lie on vertices), red represent $\sigma$ (which lie on the spatial plaquettes in the PGT, but on spatial links in its dual), and green lines represent the auxiliary spin $\lambda$ (which lie on the links along the imaginary time direction).
        Non-equal time operators are shown projected to a 2+1D subspace, with the time direction pointing ``up'' in the page.
        The three possible cut orientations are labeled by $a$,$b$, and $c$.
    }\label{fig:loops}
\end{figure}

Proceeding by analogy with the IGT, we construct the Wilson loops for the PGT and its dual  (Fig.~\ref{fig:loops}).  
Spatial loops are constructed  by choosing a set of cubes $c$ whose centers lie in a plane and taking the product of their `matchbox' terms  (terms multiplying ${K}$ in the action) such that the vacant squares of each matchbox lie parallel to the plane, resulting in a `ribbon loop' encircling it.
This can equivalently can be thought of as the dynamical process of moving a two-dimensionally mobile combination of charges around in a loop lying in a plane, via applications of the term multiplying ${J}$ in the action.
For the PGT, this is a pair of fractons, while for the dual it is a pair of parallel-moving lineons. 
Temporal Wilson loops are constructed in a similar fashion, by taking the product of the six-spin terms (that multiply ${\Gamma}$)  corresponding to each space-time cube in an $L\times L_\tau$ spacetime sheet, leaving open spatial ribbons at the initial and final slices, whose corners are linked by strings of $\lambda$s. 
This can equivalently be constructed by moving a \emph{one}-dimensionally mobile combination of charges a distance $L$ apart, evolving both for $L_\tau$ in imaginary time, and bringing them back together again.
The combination again consists of a fracton-pair in the PGT, but now only a single lineon in the dual. 
The corresponding horseshoes (or cut Wilson loop) operators are then obtained by cutting open the loop and terminating it with appropriate combination of $\tau$s, with three distinct possible orientations labeled $a$, $b$, and $c$ in Fig.~\ref{fig:loops}.

\textit{Diagnostic behaviors.}---
We now consider the expectation value of these operators at various points in the phase diagram.
First, note that the spatial Wilson loop alone functions as a diagnostic only in the pure gauge theory. When $J=0$, for small ${\Gamma}$, vison-pair fluctuations occur only on small length scales, so that only pairs along the perimeter of the loop will affect the expectation value.
In contrast,  flux excitations are condensed in the confined phase at large ${\Gamma}$, so that the loop now exhibits an area law. As in the IGT, for any $J>0$ the loop obeys a perimeter law in both phases.

Next, notice also that the spatial horseshoe alone serves as a diagnostic only along the ${\Gamma}=0$ axis, where it can be understood as measuring the vanishing of a macroscopic string tension.
To understand why this expectation value is nonzero in the Higgs/confined phase,  we draw on known results for the CPIM~\cite{johnston1}.
Early work on the ``fuki-nuke'' model~\cite{fukinuke1}, which may be thought of as an anisotropic limit of the CPIM % plaquette Ising model 
with $J=0$ for the plaquettes in the $xy$ plane, reveals that this model maps on to a stack of decoupled  2D ($xy$-planar) Ising models.
In terms of the original spins, the local observable  $\langle \tau^z_{s}\tau^z_{s+\hat{z}}\rangle$ gains a nonzero expectation value in the ordered phase, but is free to spontaneously break the symmetry in different directions for each $xy$ plane.
Now, the horseshoe operator (a) obtained by cutting open a $xy$ Wilson loop 
is exactly the correlation function of this observable:
$\langle \tau^z_{s}\tau^z_{s+\hat{z}}\tau^z_{s^\prime}\tau^z_{s^\prime+\hat{z}}\rangle$
for $s$,$s^\prime$ which are constrained to be in the same $xy$ plane, which therefore approaches a constant as $|s-s^\prime|\rightarrow\infty$ in the ordered phase.
This correlator continues to function as a diagnostic even for the isotropic model, where we are free to choose planes oriented in any direction~\cite{johnston3,fukinuke2,fukinuke3}.

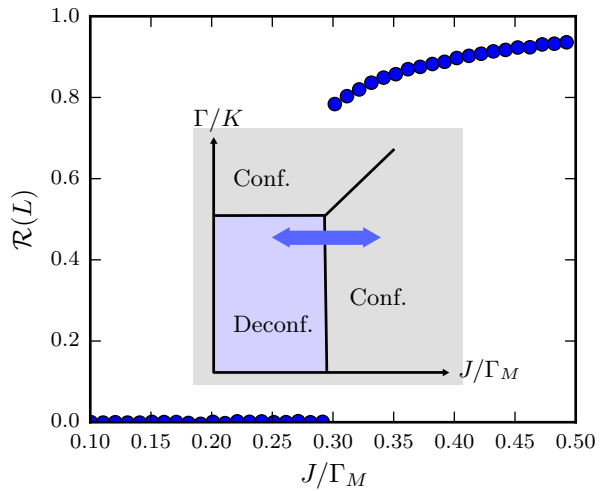
\begin{figure}[t]
    \centering
    \begin{picture}(230,200.0)%
        %\put(-10,0){\includegraphics[width=0.50\textwidth,page=1]{R.eps}}%
        %\put(75,045){\includegraphics[width=0.2\textwidth,page=1]{pd.eps}}%
        \put(-10,0){\includegraphics[width=0.50\textwidth,page=1]{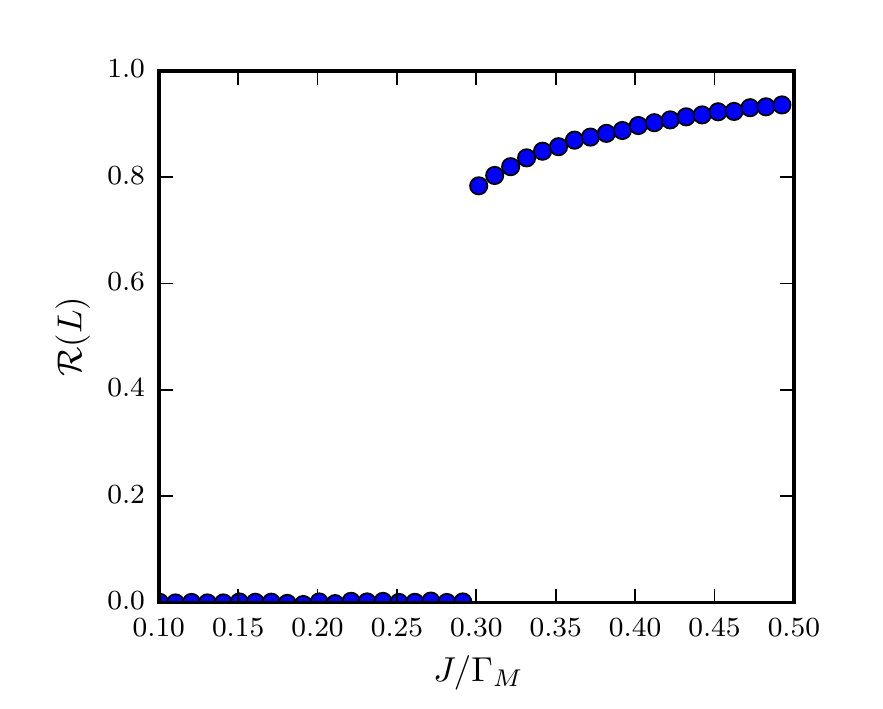}}%
        \put(75,045){\includegraphics[width=0.2\textwidth,page=1]{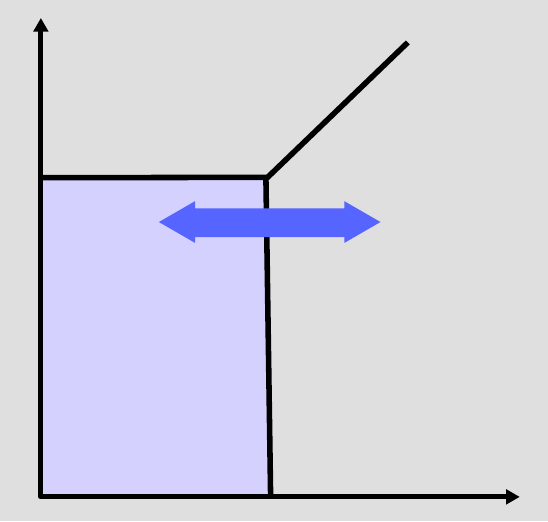}}%
        \put(175,48){\color[rgb]{0,0,0}\makebox(0,0)[lb]{\smash{$J/\Gamma_M$}}}%
        \put(75,143){\color[rgb]{0,0,0}\makebox(0,0)[lb]{\smash{$\Gamma/K$}}}%
        \put(90,65){\color[rgb]{0,0,0}\makebox(0,0)[lb]{\smash{Deconf.}}}%
        \put(134,75){\color[rgb]{0,0,0}\makebox(0,0)[lb]{\smash{Conf.}}}%
        \put(90,120){\color[rgb]{0,0,0}\makebox(0,0)[lb]{\smash{Conf.}}}%
    \end{picture}%
    \caption{The behavior of the spatial cut (Fig~\ref{fig:loops}a) ratio $\mathcal{R}(L)$ (Eq~\ref{eq:ratio}) for large $L$ across the (first-order) confinement transition as $J/\Gamma_M$ is increased with fixed $\Gamma/K=0.8$.
    Inset is a sketch of the zero-temperature phase diagram, where lines indicate first order transitions, as obtained by quantum Monte Carlo (see supplementary material~\cite{supmat} for details).
}
    \label{fig:Rpd}
\end{figure}

Away from the $J=0$ or ${\Gamma}=0$ cases, we must rely on the ratios $\mathcal{R}(L)$ (Eq.~\eqref{eq:ratio}) to distinguish between the confined and  (partially) deconfined phases.
The ratio for the spatial cut ($a$ in Fig.~\ref{fig:loops}) as before measures of the cost of opening up a gap in the loop, which depends exponentially on the size of the gap in the deconfined phase, but not in the confined phase.
Figure~\ref{fig:Rpd} shows numerical results for $\mathcal{R}(L)$ across the transition at a generic point in the phase diagram, obtained via quantum Monte Carlo calculations.
%In the supplementary material~\cite{supmat}, we verify numerically using quantum Monte Carlo that $\mathcal{R}(L)$ shows the expected behavior crossing the transition at a generic point in the phase diagram.
At ${\Gamma}=0$, $\mathcal{R}(L)$ reduces to the ``fuki-nuke'' correlation function above.

Next, we examine the temporal loops.
Consider the cut $b$ of the PGT,
$W_{1/2} = \tau^z_s\tau^z_{s+u}\tau^z_{s^\prime}\tau^z_{s^\prime+u}\prod_{p\in C^{u}_{s s^\prime}}\sigma^z_p (-T/2)$,
where $s$,$s^\prime$ are two sites on the same plane orthogonal to $u=\hat{x},\hat{y},\hat{z}$, and $C^{u}_{l l^\prime}$ defines the set of plaquettes forming a path between them (as in Figure~\ref{fig:loops}).
We have also defined $\sigma^z(T) = e^{\mathcal{H} T} \sigma^z e^{-\mathcal{H} T}$, and $T=L/c$ for a velocity $c$ in the continuum time limit $\epsilon\rightarrow0$.
Calling   our candidate two-fracton-pair  (4 fractons in total) state $|\chi\rangle = W_{1/2}|G\rangle$, created from the ground state $|G\rangle$, we see that
    $\mathcal{R}(L) = \langle G | \chi \rangle /\sqrt{\langle \chi | \chi \rangle}$
measures the overlap between the ground state and our candidate state.
This is a generalization of the Fredenhagen-Marcu diagnostic~\cite{fredenhagenmarcu1,fredenhagenmarcu2} measuring the deconfinement of fracton-pairs, with the constraint that the two fracton-pairs must be in the same plane of movement.
The final orientation of the horseshoe (cut $c$) probes the existence of delocalized fracton-pair states in the spectrum, in exactly the same way as the delocalized spinons are probed the IGT~\cite{gregor}.

Thus, rather than measuring the deconfinement of single spinons as in the IGT, our Wilson loop and horseshoe generalizations instead measure the same quantities but for the smallest mobile combinations of quasiparticles in their subspace of allowed movement.
For the PGT, this is a fracton-pair.
As stated, these diagnostics only probe the deconfinement properties of fracton-\emph{pairs}, and not single fractons.
To identify the deconfinement of individual fractons one can do the same calculation but using Wilson loops and horseshoes with a finite width that also scale with $L$.
This distinction can be important, for example, in an anisotropic version of the PGT~\cite{supmat} which exhibits
 an intermediate phase in which single fractons are confined into pairs, 
 while pairs remain deconfined (reminiscent of quark confinement into mesons).

\textit{Concluding Remarks.---}
We have shown that deconfinement diagnostics for the Ising gauge theory (or conventional topological order) can be generalized to the plaquette Ising gauge theory, which exhibits the X-cube fracton topological order in its deconfined phase.
Despite never being fully deconfined in the sense of having excitations free to move in all three dimensions,
 the expectation value of our generalized Wilson loops and horseshoes diagnoses the {\it partial} deconfinement of these excitations, with various physical interpretations depending on their orientation in Euclidean space-time. 
The procedure for identifying Wilson loop type operators is quite general, and can be extended to other similar type-I fracton models, such as the checkerboard model~\cite{vijay}. However, the extension to type-II fracton theories where the fractons (and their composites) are fully immobile remains an open question worthy of future study.

\begin{acknowledgments}
We thank Rahul Nandkishore for discussions and comments on the draft. 
This work was supported by DOE Grant No.~\uppercase{DE-SC}/0016244 (SLS). 
\end{acknowledgments}

%\end{document}

%%        File: fractonfm.tex
%%     Created: Thu Sep 14 02:00 PM 2017 E
%% Last Change: Thu Sep 14 02:00 PM 2017 E
%%
%\documentclass[twocolumn,prl, superscriptaddress]{revtex4}
%\usepackage{amsmath,graphicx,amsfonts,xcolor,float}

%\newcommand{\SP}[1]{{\color{blue} #1}}
%\newcommand{\TD}[1]{{\color{red} #1}}
%%\usepackage{times}
%\usepackage{soul}

%\begin{document}
%\title{Supplementary material for ``Correlation function diagnostics for type-I fracton phases''}
%\author{Trithep Devakul}
%\affiliation{Department of Physics, Princeton University, Princeton NJ 08540, USA}
%\author{S. A. Parameswaran}
%\altaffiliation{On leave from: Department of Physics and Astronomy, University of California Irvine, Irvine CA 92617, USA.}
%\affiliation{The Rudolf Peierls Centre for Theoretical Physics, University of Oxford, Oxford OX1 3NP, UK}
%\author{S. L. Sondhi}
%\affiliation{Department of Physics, Princeton University, Princeton NJ 08540, USA}

%\maketitle

\pagebreak
    \onecolumngrid
    \section*{\large Supplementary Material}
    \section{Path integral formulation of the plaquette Ising gauge theory}\label{app:pathintegral}
    Here, we present the derivation of the 4D Euclidean action 
    %Eq.~(\ref{eq:pgtaction})
    in terms of the auxiliary spins $\lambda_s^{(t)}$.  
    The PGT Hamiltonian is given by
    \begin{eqnarray}
        \mathcal{H}_{\text{PGT}} &=& -K\sum_{c,i}\prod_{p\in b_i(c)} \sigma^z_p - \Gamma_M\sum_{s}\tau^x_s  - J\sum_{p}\sigma^z_p\prod_{s\in\partial p}\tau^z_s- \Gamma\sum_{p}\sigma^x_p 
        \label{eq:app_pgt}
    \end{eqnarray}
    in conjunction with the constraints
    \begin{equation}
    G_s|\psi\rangle=|\psi\rangle  \hspace{0.2cm}\forall s, \hspace{0.5cm}G_s = \tau^x_s\prod_{p\in\partial s}\sigma^x_p
    \label{eq:app_pgtgauge}
    \end{equation}
    that must be satisfied on every site.

    We are interested in calculating the partition function $Z(\beta)=\text{Tr} e^{-\beta \mathcal{H}}$ for inverse temperature $\beta$ (we take $\beta\rightarrow\infty$ to access the relevant, zero-temperature limit).
    To do this, we employ the usual Suzuki-Trotter decomposition: we divide the interval $\beta$ into $L_t$ small steps of size $\epsilon$, such that $\beta=L_t\epsilon$.
    This then allows us to write the partition function as a path integral in the $z$-basis.  
    Finally, to enforce the constraint, we insert the projector into the gauge-invariant subspace at every time step, $\mathcal{P}=\prod_{s} (1+G_s)/2$.
    So, we have
    \begin{eqnarray}
        Z(\beta) &=&  \sum_{\{ \sigma^{z(t)} ,\tau^{z(t)} \}}
        \prod_{t=1}^{L_t} \langle\{ \sigma^{z(t+1)} ,\tau^{z(t+1)} \}|\mathcal{P}e^{-\epsilon\mathcal{H}}|\{ \sigma^{z(t)} ,\tau^{z(t)} \}\rangle\\\
        &=&  
        \lim_{\epsilon\rightarrow 0}
        \sum_{\{ \sigma^{z(t)} ,\tau^{z(t)} \}}
        \prod_{t=1}^{L_t} \langle\{ \sigma^{z(t+1)} ,\tau^{z(t+1)} \}|\mathcal{P}e^{-\epsilon\mathcal{H}_x}e^{-\epsilon\mathcal{H}_z}|\{ \sigma^{z(t)} ,\tau^{z(t)} \}\rangle
        \label{eq:app_z}
    \end{eqnarray}
    where in the second step we have performed a Trotter decomposition $e^{-\epsilon\mathcal{H}}\approx e^{-\epsilon\mathcal{H}_x}e^{-\epsilon\mathcal{H}_z}+\mathcal{O}(\epsilon^2)$, separating the parts of the Hamiltonian~\ref{eq:app_pgt} containing $\sigma^x,\tau^x$ and $\sigma^z,\tau^z$ into $\mathcal{H}_x$ and $\mathcal{H}_z$ respectively.
    We have also enforced periodic boundary conditions on the time direction.

    Let us now focus on evaluating a single one of these terms in the product~\ref{eq:app_z}.
    The path integral is performed in the $z$-basis, thus we can move the state past $e^{-\epsilon\mathcal{H}_z}$, picking up only a number $e^{-\epsilon\mathcal{H}_z(\{\sigma^z,\tau^z\})}$ where $\mathcal{H}_z(\{\sigma^z,\tau^z\})$ denotes $\langle\{\sigma^z,\tau^z\}|\mathcal{H}_z|\{\sigma^z,\tau^z\}\rangle$.
    Then, what's left is to compute $\langle\{ \sigma^{z\prime} ,\tau^{z\prime} \}|\mathcal{P}e^{-\epsilon\mathcal{H}_x}|\{ \sigma^{z} ,\tau^{z} \}\rangle$.

    For ease of notation, let us define the projector for $\mathcal{O}$, $P_{\mathcal{O}} \equiv (1-\mathcal{O})/2$.
    In terms of these operators, we have the following:
    \begin{eqnarray}
        \langle\{ \sigma^{x} ,\tau^{x} \}|\{ \sigma^{z} ,\tau^{z} \}\rangle &=& e^{i\pi (\sum_{p} P_{\sigma^x_p}P_{\sigma^z_p} + \sum_{s} P_{\tau^x_s}P_{\tau^z_s})} \label{eq:app_rel1}
        \\
        e^{-\epsilon\mathcal{H}_x} &\propto&  e^{-2\epsilon\Gamma\sum_p P_{\sigma^x_p} - 2\epsilon\Gamma_M\sum_s P_{\tau^x_s}}
        \label{eq:app_rel2}
    \end{eqnarray}
    where we are ignoring an overall shift in $\mathcal{H}_x$, and finally
    \begin{eqnarray}
        \mathcal{P} = \frac{1}{2^{N_s/2+N_p/2}}\prod_{s} (1 + \tau^x_{s} \prod_{p\in\partial s}\sigma^x_p) = \frac{1}{2^{N_s}}\sum_{\{\lambda_s=\pm 1\}} e^{i\pi \sum_{s} P_{\lambda_s} (P_{\tau^x_s} + \sum_{p\in\partial s} \sigma^x_p)}
        \label{eq:app_rel3}
    \end{eqnarray}
    where $N_s$ ($N_p$) is the number of sites (plaquettes), and we introduced an Ising variable $\lambda_s$ to mediate the constraint on site $s$.

    Inserting a resolution of the identity $\mathbf{1} =\sum_{\{\sigma^x,\tau^x\}} |\{\sigma^x,\tau^x\}\rangle\langle\{\sigma^x,\tau^x\}|$, and using Eqs.~(\ref{eq:app_rel1}-\ref{eq:app_rel3}), we get
    \begin{eqnarray}
\langle\{ \sigma^{z\prime} ,\tau^{z\prime} \}|\mathcal{P}e^{-\epsilon\mathcal{H}_x}|\{ \sigma^{z} ,\tau^{z} \}\rangle
= 
\frac{1}{2^{2N_s+N_p}}\sum_{\{\lambda_s\}}\sum_{\{\sigma^x_p,\tau^x_s\}} &&e^{\sum_p P_{\sigma^x_p} (-2\epsilon\Gamma+i\pi[P_{\sigma^z_p} + P_{\sigma^{z\prime}_p}+\sum_{s\in\partial p} P_{\lambda_s} ])}\\
&&\times e^{\sum_s P_{\tau^x_s} (-2\epsilon\Gamma_M + i\pi[P_{\tau^{z\prime}_s}+P_{\tau^z_s}+P_{\lambda_s}])}
        \label{}
    \end{eqnarray}
    %Performing the sum over $\{\sigma^x_p,\tau^x_s\}$,
    \begin{eqnarray}
 %\langle\{ \sigma^{z\prime} ,\tau^{z\prime} \}|\mathcal{P}e^{-\epsilon\mathcal{H}_x}|\{ \sigma^{z} ,\tau^{z} \}\rangle       
 &=& 
 \frac{1}{2^{2 N_s+N_p}}\sum_{\{\lambda_s\}} \prod_p (1+e^{-2\epsilon\Gamma+i\pi[P_{\sigma^z_p} + P_{\sigma^{z\prime}_p}+\sum_{s\in\partial p} P_{\lambda_s} ]})
 \prod_{s} (1+e^{-2\epsilon\Gamma_M + i\pi[P_{\tau^{z\prime}_s}+P_{\tau^z_s}+P_{\lambda_s}]})\\
 &=& 
 \frac{1}{2^{2 N_s+N_p}}\sum_{\{\lambda_s\}} \prod_p (1+e^{-2\epsilon\Gamma}\sigma^z_p\sigma^{z\prime}_p\prod_{s\in\partial p} \lambda_s )
 \prod_{s} (1+e^{-2\epsilon\Gamma_M}\tau^{z\prime}_s\tau^z_s\lambda_s)\\
 &\propto& \sum_{\{\lambda_s\}} e^{\tilde{\Gamma}\sum_p \sigma^{z\prime}_p \sigma^z_p \prod_{s\in\partial p}\lambda_s + \tilde{\Gamma}_M \sum_s \tau^{z\prime}_s\tau^z_s\lambda_s}
        \label{}
    \end{eqnarray}
    where $\tilde{\Gamma} = -\frac{1}{2}\log\tanh\epsilon\Gamma$ and $\tilde{\Gamma}_M = -\frac{1}{2}\log\tanh\epsilon\Gamma_M$.
    Thus, $\lambda_s$ can be thought of as a spin variable %degree of freedom 
    located on the bond between site $s$ at time $t$ and $t+1$.  
    Labelling each $\lambda_s^{(t)}$ by the time index and combining all our parts, the total partition function is given by $Z(\beta) \propto \sum_{\{\sigma^{(t)},\tau^{(t)},\lambda^{(t)}\}} e^{-
        \mathcal{S}_{\text{PGT}}(\{\sigma^{(t)},\tau^{(t)},\lambda^{(t)}\})}$ where we have suppressed the $z$ label on $\sigma^{(t)},\tau^{(t)}$, with the action
    \begin{eqnarray}
        \mathcal{S}_{\text{PGT}}
    &=& -\tilde{K}\sum_{t,c,i}\prod_{p\in b_i(c)} \sigma_p^{(t)} - \tilde{\Gamma}_M\sum_{t,s}\tau_s^{(t)}\tau_s^{(t+1)}\lambda_s^{(t)} 
     -\tilde{J}\sum_{t,p}\sigma^{(t)}_p\prod_{s\in\partial p}\tau_s^{(t)} - \tilde{\Gamma}\sum_{t,p}\sigma_p^{(t)}\sigma_p^{(t+1)}\prod_{s\in\partial p}\lambda_s^{(t)} 
    \end{eqnarray}
    where we have defined $\tilde{K}=\epsilon K$ and $\tilde{J}=\epsilon J$.
    Note that the gauge constraint manifests as a local symmetry in the action: a simultaneous flip of $\tau_s^{(t)}$, $\sigma_p^{(t)}$ for $p\in\partial s$,  $\lambda_s^{(t)}$, and $\lambda_s^{(t-1)}$ leaves the action unchanged.
    Thus, we have successfully obtained the Euclidean action for the PGT\@.  The zero temperature limit can be taken by making the time direction infinite.

    Finally, 
    due to the Ising nature of these variables, we may now express the partition function as a sum of products involving every possible combination of terms in the action,
    \begin{eqnarray}
        Z \propto \text{Tr}_{\{\sigma,\tau,\lambda\}}
        &&\prod_{t,c,i}\left(1+[\tanh \tilde{K}]\prod_{p\in b_i(c)}\sigma_p^{(t)}\right)\prod_{t,s}\left(1+[\tanh\tilde{\Gamma}_M]\tau_s^{(t)}\tau_s^{(t+1)}\lambda_s^{(t)}\right)\\
        &&\times
        \prod_{t,p}\left( 1+[\tanh \tilde{J}] \sigma_p^{(t)}\prod_{s\in\partial p}\tau_s^{(t)} \right) \prod_{t,p} \left( 1+[\tanh \tilde{\Gamma}]\sigma_p^{(t)}\sigma_p^{(t+1)}\prod_{s\in\partial p} \lambda_s^{(t)}  \right)
        \label{}
    \end{eqnarray}
    Expanding the product, any term that contains an odd number of any $\sigma_p^{(t)},\tau_s^{(t)},\lambda_{s}^{(t)}$ vanish under the trace.
    Thus, only combinations in which each of these appear an even number of times contribute to the partition function.
    This can therefore be thought of as a statistical mechanical model of edges, and surfaces, where each configuration appears with its own weights, but with  a more complex set of rules for allowed shapes than in the edge-surface statistical mechanical interpretation that can be given to the Euclideanized partition function of a  conventional gauge theory.
    Nevertheless, it is still possible to assign an interpretation of the confinement/deconfinement transition in terms of vanishing string and surface tensions: $\tilde{K}$ ($\tilde{\Gamma}$) play the role of a surface cost in the space (time) directions, and $\tilde{J}$ ($\tilde{\Gamma}_M$) play the role of the edge cost in the space (time) directions.
    In this language, the deconfined phase corrsponds to a phase with zero (macroscopic) surface tension and high line tension, and the confined phase to one where either surface tension is nonzero or line tension is zero.

\section{The anisotropic plaquette Ising gauge theory}\label{app:fukinuke}
    Let us consider the anisotropic plaquette Ising gauge theory.  
    To simplify notation, let us redefine $\sigma^z_p\prod_{s\in\partial p}\tau_s^{z}\rightarrow \sigma^z_p$, and use the constraint to eliminate $\tau_s^{x} = \prod_{p\in\partial s}\sigma^x_p$, so that the Hamiltonian can be expressed without reference to the $\tau$ spins.
    We wish to couple the plaquettes in the $xy$-plane with a weaker coupling $J^\prime$, and consider the resulting gauge theory.  
    When $J^\prime=0$, the classical plaquette model coincides with the fuki-nuke model~\cite{app-fukinuke1}, which decouples into a stack of 2D Ising models.
    For simplicity, we take the limit ${\Gamma}_M\gg J,J^\prime$, which allows us to effectively set $J=J^\prime=0$ and only look at their effects perturbatively in the form of the ${K}$,$K^\prime$ terms.
    The anisotropic Hamiltonian is defined by
    \begin{eqnarray}
        \mathcal{H}_{\text{aPGT}} &=& -K\sum_{c}\prod_{p\in b_1(c)} \sigma^z_p - K^{\prime}\sum_{c,i\in\{2,3\}}\prod_{p\in b_i(c)} \sigma^z_p 
         - \Gamma\sum_{p}\sigma^x_p 
- \Gamma_M\sum_{s}\prod_{p\in\partial s}\sigma^x_p
        \label{eq:app_apgt}
    \end{eqnarray}
    where $b_1(c)$ is the matchbox operator that does not contain the $xy$-plaquettes.
    The gauge theory that one would have obtained from the fuki-nuke model is obtained from this Hamiltonian by setting $K^\prime=0$ and fixing $\sigma^x_p=1$ for the $xy$-plaquettes
    (notice that we could have taken advantage of the fact that the fuki-nuke model could be written as a stack of decoupled Ising models and done the gauging process with those spins instead, in which case its corresponding gauge theory would have been trivially a stack of decoupled 2D Ising gauge theories.)

    The goal of this section will be to show using perturbative arguments that this anisotropic model possesses deconfined fracton excitations when ${\Gamma}\ll K^\prime\ll K$, but that these fracton excitations become confined into fracton-pairs (which are themselves still deconfined) when $K^\prime\ll\Gamma\ll K$, and that these pairs too eventually become confined at ${\Gamma}\gg K$.

    Note that the first limit is exactly the X-cube model with a small perturbation ${\Gamma}\sigma^x$, which does not confine the fractons.
    To leading order in the effective Hamiltonian, this perturbation produces a term of the form $\prod_{p\in\partial s}\sigma^x_p$, which is exactly the ${\Gamma}_M$ term.  
    Similarly, the last limit is just the large ${\Gamma}$ limit of the X-cube model.  
    All $\sigma^x$ are fixed to $+1$ with energy gap ${\Gamma}$, and to create two fracton-pairs requires flipping $\sigma^x$ proportional to their separation.  
    The cost of separating two pairs therefore scales with their distance, and so all such excitations are confined.

    We now focus on the intermediate limit, $K\gg \Gamma \gg K^\prime$.  
    Notice that the term ${\Gamma}\sigma^x_p$ commutes with the ${K}$ term in the Hamiltonian if $p\perp z$ ($p$ is an $xy$-plaquette).  
    Thus,  we can write
    \begin{eqnarray}
        \mathcal{H}_{\text{aPGT}} &=& -K\sum_{c}\prod_{p\in b_1(c)} \sigma^z_p - \Gamma\sum_{p\perp z}\sigma^x_p - \Gamma_M\sum_{s}\prod_{p\in\partial s}\sigma^x_p 
        + \left[ -K^{\prime}\sum_{c,i\in\{2,3\}}\prod_{p\in b_i(c)} \sigma^z_p  
        - \Gamma\sum_{p\not\perp z}\sigma^x_p \right]
        \label{eq:app_apgt2}
    \end{eqnarray}
    and proceed perturbatively in the terms in the square bracket.
    The $K^\prime$ term contributes to the effective Hamiltonian (in the ground state manifold) in leading order as $\prod_{p\in b_1(c)}\sigma^z_p$, which is exactly the first term, and thus does not change anything.
    However, the ${\Gamma}$ term results in the 4th order in terms of the form $-\Gamma_{\text{eff}}\sum_{ l \parallel z}\prod_{p\in\partial l } \sigma^x_p$ for each link $l$ pointing in the $z$ direction, where the product is over the four plaquettes that touch it (and we do not care about the exact value of the coefficient $\Gamma_\text{eff}>0$).
    This term comes with a minus sign since the three virtual steps in the perturbation theory always involve states higher in energy.

    What is the unperturbed ground state?  The first three terms all mutually commute, so we can satisfy all of them simultaneously.  
    Working in the $\sigma^x$ basis, we see that the ${\Gamma}$ term means that for all $ p\perp z, \sigma^x_p=1$.
    In combination with this, the ${\Gamma}_M$ term means that  for each link $l\parallel z$, we must have $\prod_{p\in\partial l}\sigma^x_p = \prod_{p\in\partial l\pm z}\sigma^x_p$, where $l\pm z$ corresponds the link directly above or below $l$ in the $z$ direction.
    Thus, for the column located at $(x,y)$, this product $\prod_{p\in\partial l}\sigma^x_p$ must either be $+1$ or $-1$ for the entire column.
    Finally, satisfying the ${K}$ term means that the ground state is an equal superposition of all possible configurations of $\sigma^x_p=-1$ that is reachable by repeated applications.

    In addition to the topological degeneracy corresponding to the winding number within each $z$-layer, there is an additional large degeneracy from picking the possible configurations of columns to be $\pm 1$.  
    This arises from an extra symmetry where one is free to flip all $\tau^z$ along any column, and is only present in the completely anisotropic limit.  
    This degeneracy is broken as soon as ${\Gamma}>0$, which penalizes having a $-1$ column.  
    Thus, we will work with every column being $+1$.  

    Having figured out the ground state, we can construct the exact four-fracton eigenstate by acting with the membrane operator $\prod_{p\in \Sigma} \sigma^z_p$, where $\Sigma$ defines a rectangular membrane that creates four fracton (${\Gamma}_M$) excitations at its corners, which are separated by a distance $L$ in the $x$ or $y$ direction, and $L_z$ in the $z$ direction.  
    This results in two columns in which $\prod_{p\in\partial l}\sigma^x_p = -1$ for a section of length $L_z$.
    Introducing the perturbation penalizes this column with an energy proportional to $L_z$.  
    Thus, the energy of separating a fracton-pair a distance $L_z$ apart scales proportionally with $L_z$, and so single fractons are confined.

    Having shown this, there is an alternate view of the situation, in terms of quasiparticles and mutual statistics.  
    For ${\Gamma} \ll K^\prime \ll K$ we are in the X-cube regime.  
    ${\Gamma}$ creates fluctuations of one-dimensional quasiparticles.  
    These can ``braid'' with single fractons through an off-shell process in which they form a box, which gains a minus sign (and thus an energy penalty) if a single fracton is contained inside the box. 
    These boxes are small, and so a state with four well-separated fractons only cost additional energy proportional to the number of fractons which is constant and therefore deconfined.

    When $K^\prime \ll \Gamma \ll K$, the one-dimensional quasiparticle that moves in the $z$ direction ($z$-movers) become condensed.  
    This allows an $x$-mover to become a $y$-mover for free via an exchange of $z$-movers with the condensate.  
    Thus, our flux excitations are now mobile within an $xy$-plane.  
    These can circle around the column separating two fractons and gain a minus sign, and therefore costs energy proportional to their separation.
    We see, as is usual, that a condensation transition corresponds to the confinement transition of another particle with which the condensate has nontrivial statistics, which in this case corresponds to the confinement of single-fractons.  
    These fracton-pairs are also only able to move in the $xy$ plane.
    Thus, we see that this phase corresponds to the stacked $\mathbb{Z}_2$ topological order, as one would have expected from the decoupled-plane structure of the fuki-nuke limit. Note that the fuki-nuke limit is a distinct `layer construction' from the isotropic constructions of Refs.~\onlinecite{app-layered,app-layered2}.
    Finally, we note that this model in the completely anisotropic limit bears many similarities to the anisotropic fracton model presented in Ref~\onlinecite{app-petrova}, such as the large non-topological subextensive degeneracy.

\section{Numerical Verification and phase diagram}
In this section, we perform some modest numerical calculations for the PGT using quantum Monte Carlo (QMC) simulations.
We perform these simulations using the stochastic series expansion (SSE) formalism~\cite{app-SSE} for simplicity.
For the purpose of the calculation, we gauge fix $\tau^z=1$ and move to the dual lattice.
In the dual lattice, $\sigma$ degrees of freedom live on the links $l$.
The Hamiltonian $\mathcal{H}_\text{PGT}$ describes the X-cube model with $\sigma^x$ and $\sigma^z$ perturbations,
\begin{equation}
\mathcal{H}_{\text{PGT}} = -K\sum_{x} \prod_{l\in \partial x}\sigma^z_l - \Gamma_M\sum_{c}\prod_{l\in \partial c} \sigma^x_l 
- J\sum_{l}\sigma^z_l - \Gamma\sum_{l}\sigma^x_l 
\end{equation}
where $x$ represents crosses (of which there are three per vertex), $c$ represents all cubes, $l\in \partial x$ represents the four links taking part in the cross, and $l\in\partial c$ the 12 links along the edges of the cube.  We also assume all parameters are positive.

\subsection{Stochastic series expansion}

For the purpose of the calculation, we introduce the operators 
\begin{eqnarray}
    H_{0,0} &=& 1 \label{eq:app_beginops}\\
    H_{l,0} &=& C_{l} + J\sigma_l^z \\
    H_{l,1} &=& \Gamma\sigma_l^x \\
    H_{c,0} &=& C_{c} \\
    H_{c,1} &=& \Gamma_M \prod_{l\in\partial c} \sigma_l^x \\
    H_{x,0} &=& C_{x} + K\prod_{l\in\partial x} \sigma_l^z \label{eq:app_endops}
\end{eqnarray}
for each link $l$, cube $c$, and cross $x$, such that 
\begin{equation}
    \mathcal{H}_\text{PGT} = -\sum_{l,j} H_{l,j} - \sum_{c,j} H_{c,j} - \sum_{x,j} H_{x,j}
\end{equation}
up to a constant (and $j=0,1$ represents diagonal or offdiagonal terms, in the $\sigma^z$-basis).
The constants $C_{l}$, $C_{c}$, and $C_{x}$ must be chosen such that all these terms are positive.
Here, we choose $C_l = \max(J,\Gamma)+0.5$, $C_c = \Gamma_M$, and $C_x=K+0.5$.

In the SSE approach~\cite{app-SSE}, we expand the partition function
\begin{eqnarray}
    Z = e^{-\beta \mathcal{H}_\text{PGT}} &=& \sum_{\alpha} \sum_{n=0}^{\infty} \frac{\beta^n}{n!}\langle \alpha | {(-\mathcal{H}_\text{PGT})}^n | \alpha \rangle\\
    &=& \sum_{\alpha} \sum_{n=0}^{\infty} \sum_{S_n} \frac{\beta^n}{n!}\langle \alpha | \prod_{i=1}^{n} H_{s(i),j(i)} | \alpha \rangle
\end{eqnarray}
where $S_n$ designates a particular sequence of operators by their label,
\begin{equation}
    S_n = [s(1),j(1)],[s(2),j(2)], \dots [s(n), j(n)]
\end{equation}
 $s(i)$ designates a link, cube, or cross, and $j(i)=0,1$ (except when $s(i)$ is a cross, in which case we only have $j(i)=0$).
 The sum over $\alpha$ is over all product states $|\alpha\rangle = |\left\{\sigma^z_l\right\}\rangle$.
 
 To construct an efficient sampling scheme, the expansion is truncated at some power $n=M$ sufficiently high that the cutoff error is negligible (in practice $M$ is increased dynamically as necessary until there is no cutoff error).
 A further simplification is obtained by keeping the length of the operator string $S_n$ fixed, and allowing $M-n$ unit operators $H_{0,0}$ to be present in the operator list.  
 Correcting for the $M\choose n$ possible ways that $H_{0,0}$ may be inserted in the list gives
 \begin{equation}
     Z =  \sum_{\alpha}\sum_{S_M} \frac{\beta^n (M-n)!}{M!} \langle \alpha | \prod_{i=1}^{M} H_{s(i),j(i)} | \alpha \rangle
     \label{eq:app_zsse}
 \end{equation}
where now $[s(i),j(i)] = [0,0]$ is a valid entry in $S_M$, and $n$ only counts the number of non-identity operators.
For convenience, we define the state $|\alpha(i)\rangle = |\left\{\sigma^z_l(i)\right\}\rangle$ obtained by propagating $|\alpha\rangle$ by the first $i$ operators of $S_M$.
We then need to step through the space of possible configurations $S_M$ and states $\alpha$, with probability proportional to the weight in the partition function sum Eq.~\ref{eq:app_zsse}.

\subsection{Update procedure}
Due to the four and twelve-spin interactions, along with the arbitrary transverse and longitudinal fields, naively we cannot efficiently apply non-local update techniques such as loop or cluster algorithms~\cite{app-loopclus,app-loopclus2}.
Notice that along the particular axes $J=0$ or $\Gamma=0$, the model can be mapped on to Ising models with a transverse field, for which more efficient algorithms can surely be devised.  
% However, we are interested in general points in the pahse diagram here.
Here, we apply a simple spatially local Metropolis update procedure.  
A Monte carlo step consists of a diagonal update step, followed by a number of off-diagonal updates, which we will detail below.

\subsubsection{Diagonal update}
The diagonal update consists of stepping through the $M$ elements of $S_M$.  If an off-diagonal operator $[s(i),1]$ is encountered, we continue on to the next operator in the list.  If a diagonal ($[s(i),0]_i$) or identity ($[0,0]_i$) operator is encountered, we propose to replace it with an identity or diagonal operator, respectively (the subscript $i$ indicates the position of the operator in $S_M$).  
If a diagonal operator $[s(i),0]_i$ is encountered, we remove it with probability
\begin{equation}
    P([s(i),0]_i\rightarrow[0,0]_i) = \frac{M-n+1}{\beta \left[N_l (C_l+J) + N_x (C_x+K) + N_c C_c \right]}
\end{equation}
where $N_l$, $N_x$, $N_c$ is the total number of links, crosses, and cubes.

If an identity operator is encountered, we propose to add a diagonal operator with probability
\begin{equation}
    P([0,0]_i\rightarrow[s(i),0]) = \frac{\beta \left[N_l (C_l+J) + N_x (C_x+K) + N_c C_c \right]}{M-n}
\end{equation}
If we have decided to add an operator, we must further decide the type of operator.  
The type of operator to add chosen with probabilities
\begin{eqnarray}
    P(\text{link}) &=& \frac{N_l (C_l+J)}{N_l (C_l+J) + N_x (C_x+K) + N_c C_c }\\
    P(\text{cross}) &=& \frac{N_x (C_x+K)}{N_l (C_l+J) + N_x (C_x+K) + N_c C_c }\\
    P(\text{cube}) &=& \frac{N_c C_c}{N_l (C_l+J) + N_x (C_x+K) + N_c C_c }\\
\end{eqnarray}
If the type chosen is a link, we randomly pick a link $l$ and insert a diagonal bond operator $[l,0]$ with probability 
\begin{equation}
    P(\text{add link } l) = \frac{C_l + J\sigma^z_l(i)}{C_l+J}
\end{equation}
otherwise, if the type chosen is a cross, we randomly pick a cross $x$ and insert $[x,0]$ with probability
\begin{equation}
    P(\text{add cross } x) = \frac{C_x + K\prod_{l\in \partial x}\sigma_l^z(i)}{C_x+K}
\end{equation}
and finally, if a cube is chosen, we choose a random cube $c$ and insert the operator $[c,0]$ with probability $1$.
If we fail any of these probability checks, we simply consider the move failed and continue on to the next element $i+1$ in $S_M$.
This concludes the diagonal update step.

\subsubsection{Offdiagonal update}
We perform simple local offdiagonal updates.  
These come in two types, link operator flips and cube operator flips.

The link operator flip consists of picking a link operator (diagonal or offdiagonal) $[l,j(i)]_i$ randomly in $S_M$.  
We then find the next operator acting on the same link, $[l,j(i^\prime)]_{i^\prime}$, $i\neq i^\prime$, and propose to flip the spin state between the two, which we accept with a Metropolis probability
\begin{equation}
    P\left([l,j(i)]_i [l,j(i^\prime)]_{i^\prime} \rightarrow [l,\bar{j}(i)]_i [l,\bar{j}(i^\prime)]_{i^\prime}\right) = \min\left(1, \frac{W_\text{new}}{W_\text{old}}\right)
\end{equation}
where $W_\text{new}/W_\text{old}$ is the ratio of the weights after and before the flip, and $\bar{j}(i)=1-j(i)$.
The weight difference depends only on the difference between the number of satisfied and dissatisfied cross operators acting on the link $l$ between $i$ and $i^\prime$.
Letting $n_{\pm}$ be the number of $\prod_{l\in\partial x} \sigma^z_l=\pm 1$ cross operators acting on site $l$ between $i$ and $i^\prime$ \emph{before} the flip. 
The weight ratio before and after the flip is simply given by 
\begin{equation}
    \frac{W_\text{new}}{W_\text{old}} = {\left(\frac{C_x-K}{C_x+K}\right)}^{n_+-n_-}
\end{equation}

The cube operator flip is similarly a flip of two consecutive cube operators acting on the same cube $c$.  
We randomly pick an operator $[c,j(i)]_i$ and its next $[c,j(i^\prime)]_{i^\prime}$, and propose to flip the state of all 12 spins between the two, which we again accept with probability
\begin{equation}
    P\left([c,j(i)]_i [c,j(i^\prime)]_{i^\prime} \rightarrow [c,\bar{j}(i)]_i [c,\bar{j}(i^\prime)]_{i^\prime}\right) = \min\left(1, \frac{W_\text{new}}{W_\text{old}}\right)
\end{equation}
Notice that since the cube shares two links with any cross, there is no weight change due to cross operators between $i$ and $i^\prime$.
The only weight change due to this flip comes from diagonal link operators $[l,0]$.
Similarly to before, letting $n_{\pm}$ be the number of $[l,0]$ operators acting on the involved links between $i$ and $i^\prime$ with $\sigma^z_l = \pm1$, we have the weight ratio
\begin{equation}
    \frac{W_\text{new}}{W_\text{old}} = {\left(\frac{C_l-J}{C_l+J}\right)}^{n_+-n_-}
\end{equation}

Finally, we note that including only these moves is not sufficient for ergodicity, as the total number of offdiagonal link operators acting on link $l$ is always even, and the total number of offdiagonal cube operators acting on a cube $c$ is also always even.
We can presumably make the algorithm ergodic by allowing moves in which one cube operator is flipped along with 12 link operators.  
We do not consider such moves, as the parity of such operators is a non-local measurement (along the time direction), and should be locally indistinguishable.  We have verified that including such moves do not make a discernible difference.
Also, since we have periodicity along the expansion direction, the offdiagonal flips that cross the boundary also sample through states $|\alpha\rangle$.

A full Monte carlo step then consists of the diagonal update step, followed by a number of link and cube offdiagonal updates.
We begin the system with some $M$ and $S_M$ consisting of only identity operators.
As the number of non-identity operators $n$ increases, we increase $M$ such that $M>(3/2)n$ at all times, so that the truncation error is completely negligible.

\subsection{Results}
Finally, we present modest numerical results using the above local update procedure.
We perform simulations on a $10\times 10 \times 10$ periodic lattice with $K=\Gamma_M=1$ at $\beta=8$, and consider the breakdown of the deconfined phase as we introduce $J$ and $\Gamma$.
We have verified that our choice of $\beta$ is high enough that we are essentially seeing only ground state behavior.
The present algorithm is also prone to getting ``stuck'' in suboptimal configurations, but manages to find the correct ground states in the various ($J=\Gamma=0$, large-$J$, or large-$\Gamma$) limits following a slow ramp from $\beta=0$ (infinite temperature).

\begin{figure}
    \centering
    \includegraphics[width=0.33\textwidth]{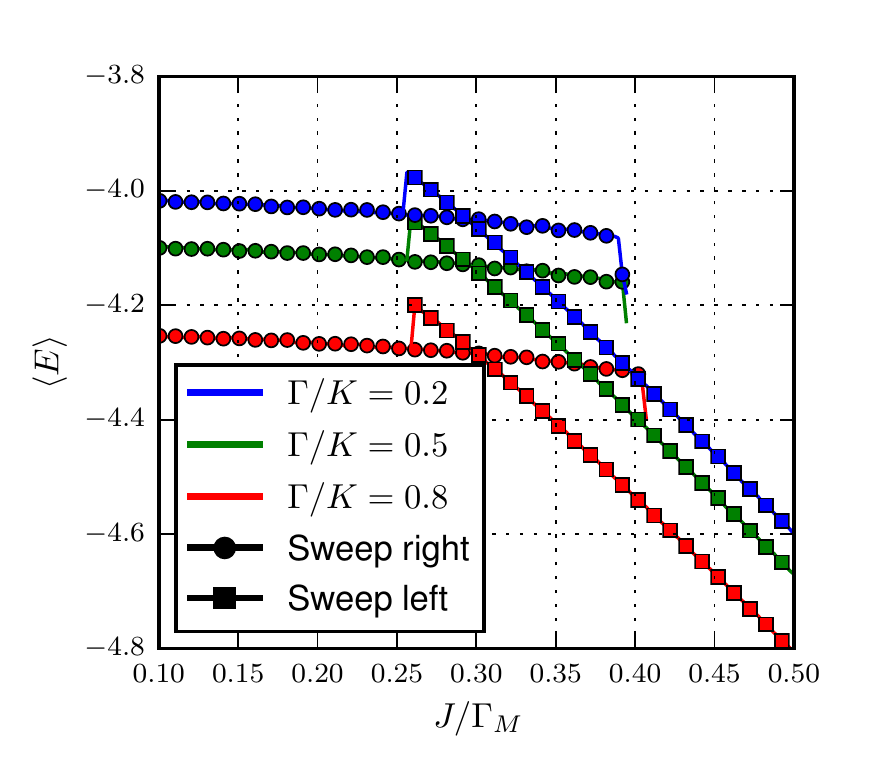}%
    \includegraphics[width=0.33\textwidth]{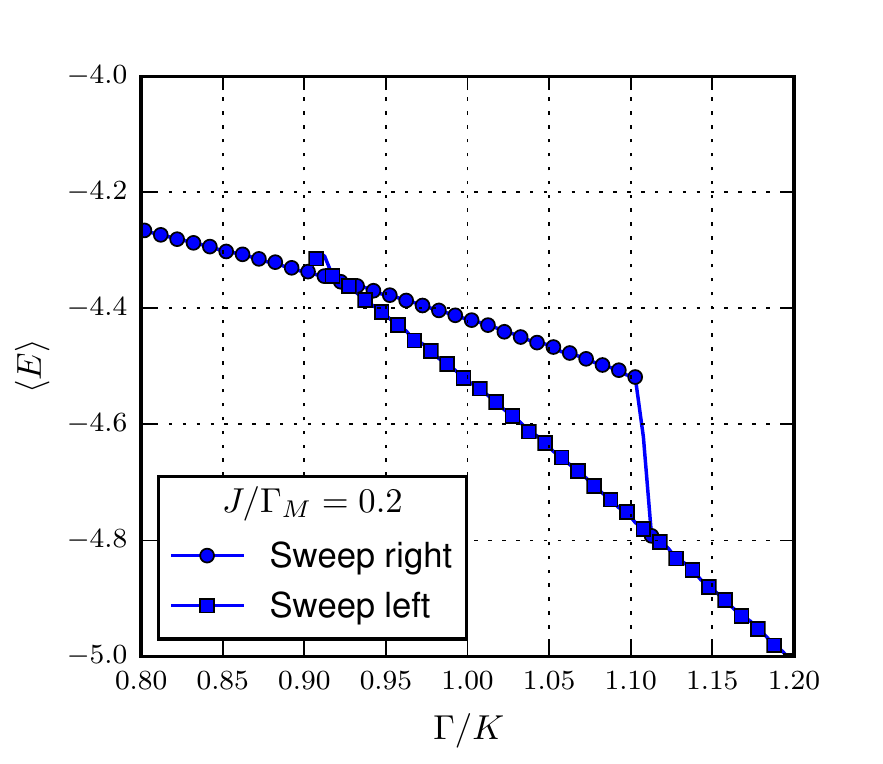}%
    \includegraphics[width=0.33\textwidth]{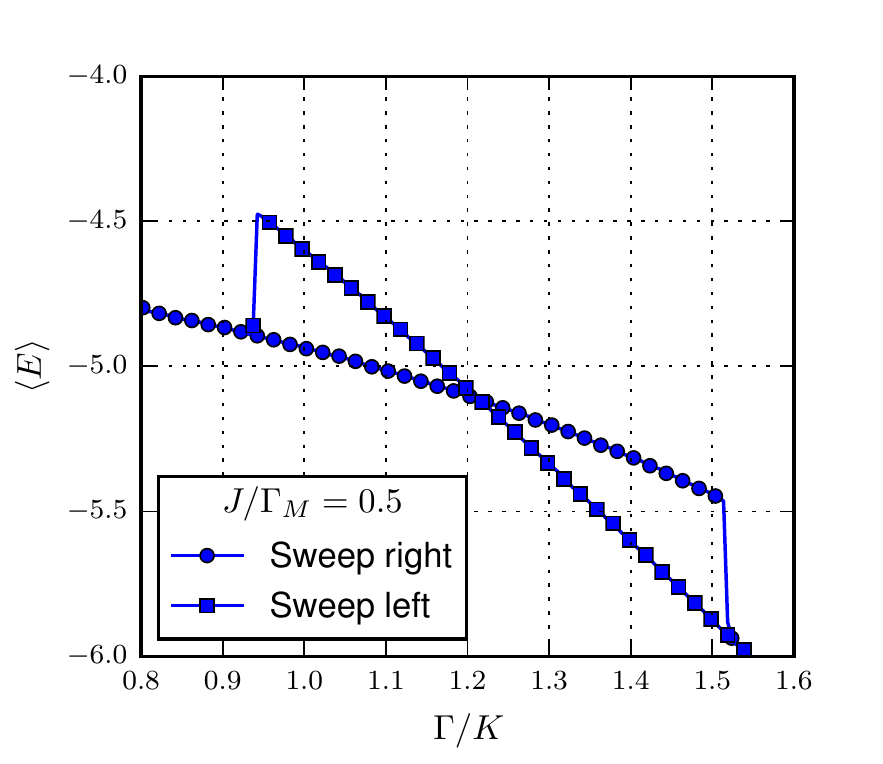}
    \caption{Plots of the energy $\langle E \rangle$ as a function of $J$ and $\Gamma$, for a $10\times 10\times 10$ system at $\beta=8$ with $K=\Gamma_M=1$, with the QMC constants (Eq~\ref{eq:app_beginops}-\ref{eq:app_endops}) subtracted out.
    We only show data until the QMC state becomes unstable and transitions into a lower-energy state.
    (left) Sweeping $J$ at various values of $\Gamma$, sweeping right from the X-cube limit and left from the trivial $\sigma^z=1$ large-$J$ limit, showing strong first order transitions at $J/\Gamma_M\approx0.3$.
    (center) Sweeping of $\Gamma$ with $J=0.2$, showing a first-order transition at around $\Gamma/K\approx0.93$ (other values of $J<0.3$ look very similar).
    (right) Sweeping $\Gamma$ with $J=0.5$ (which is confining), showing a first-order transition between the two confined phases.
    }
    \label{fig:app_transition}
\end{figure}

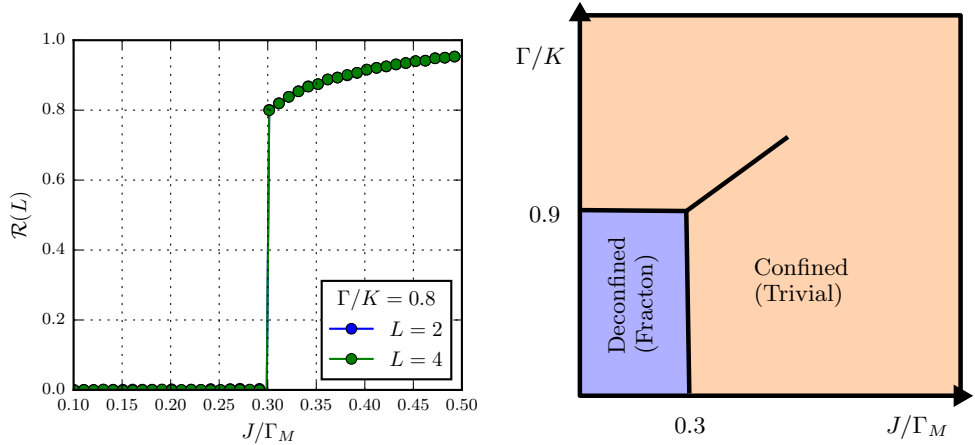
\begin{figure}
    \centering
    \includegraphics[width=0.4\textwidth]{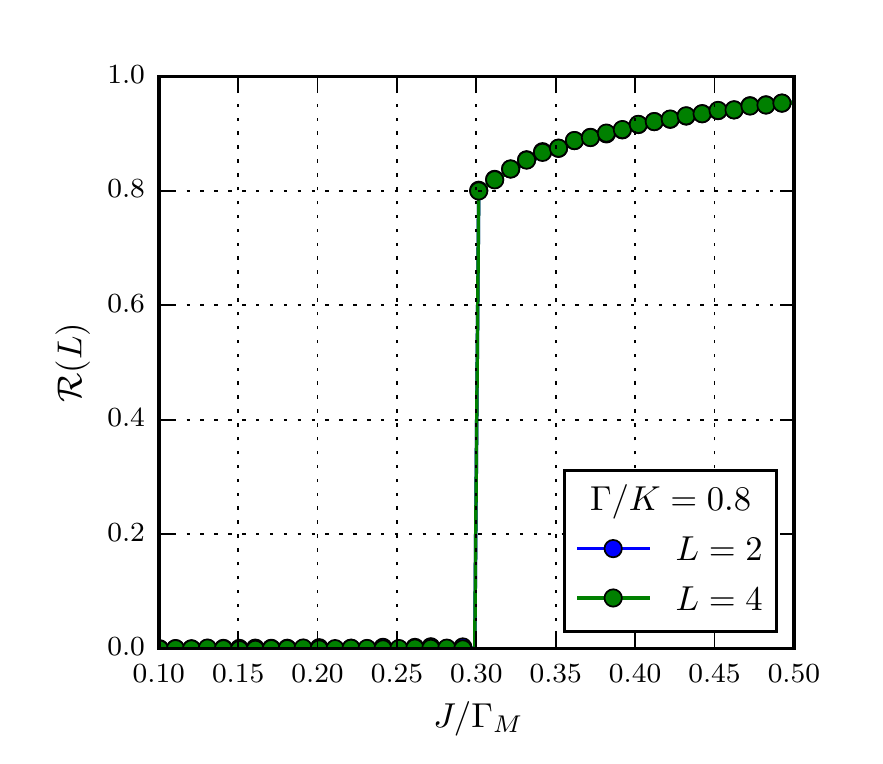}%
    \begin{picture}(150,150)
    \put(20,20){\includegraphics[width=0.3\textwidth]{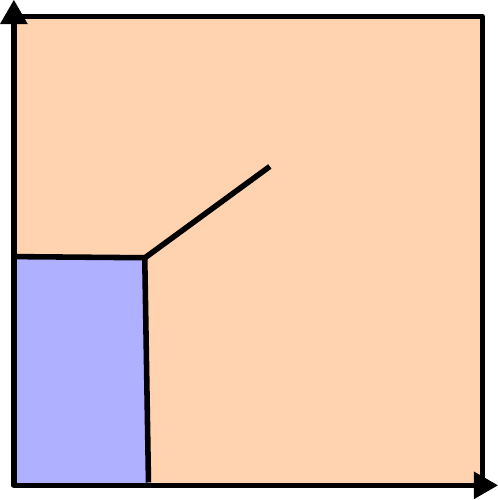}}
    \put(140,10){$J/\Gamma_M$}
    \put(00,150){$\Gamma/K$}
    \put(90,70){Confined}
    \put(90,60){(Trivial)}
    \put(35,40){\rotatebox{90}{Deconfined}}
    \put(45,40){\rotatebox{90}{(Fracton)}}
    \put(60,10){$0.3$}
    \put(5,90){$0.9$}
    \end{picture}
    \caption{(left) The expectation value of the diagnostic $\mathcal{R}(L)$ defined in the main text, which approaches zero (a constant) in the deconfined (confined) phase as $L\rightarrow\infty$.
    Here, the loop is taken to be an $L\times L$ square, and the horseshoe has dimensions $L/2\times L$.
    We look at the transition induced by increasing $J$ at fixed $\Gamma/K=0.8$.
    The correlation lengths are very short near the first order transition and already $L=2$ is indistinguishable from $L=4$, thus we are  already in the large-$L$ limit and $\mathcal{R}(L)$ shows the expected behavior.  
    Note that we only show the lower-energy state at the first order transition.
    (right) A schematic phase diagram summarizing the sweep results from Figure~\ref{fig:app_transition}.
    All transitions are first-order.
    }
    \label{fig:app_phasediagram}
\end{figure}

Figure~\ref{fig:app_transition} shows the internal energy $\langle E \rangle$ per site (with the additional constants introduced for the QMC calculation in Eq~\ref{eq:app_beginops}-\ref{eq:app_endops} subtracted out), as a function of $J$ and $\Gamma$ perturbations, where the QMC system is swept along both increasing and decreasing $J$ and $\Gamma$.
Looking at the energy allows one to identify the confinement transition, which appears to be strongly first order everywhere, as evidenced by the strong hysteresis which appears to be independent of sweeping rate.
The confinement transition occurs at roughly $J/\Gamma_M \approx 0.3$ or $\Gamma/K\approx 0.9$.
Finally, we note that akin to the phase diagram of the Ising gauge theory~\cite{app-igtpd1,app-igtpd2,app-igtpd3,app-igtpd4,app-igtpd5,app-igtpd6} there appears to be a line of first order transition extending from the corner of the deconfined phase, as shown in Figure~\ref{fig:app_transition}(right) (which are smoothly connected in the large-$J,\Gamma$ limit, where the Hamiltonian becomes simply a rotation of a field).
These result in the phase diagram shown in Figure~\ref{fig:app_phasediagram}(right).
Note that this line of first order transition must terminate at a critical point, where one can perform scaling analysis.
We leave a more complete analysis of the phase diagram to future work.

In Figure~\ref{fig:app_phasediagram}(left), we show the behavior of the diagnostic $R(L)$ introduced in the main text for length $L=2$ and $L=4$ loops, across the confinement transition as we increase $J$ keeping $\Gamma=0.8$.
These small loops are already enough for convergence, as $R(L)$ is already independent of $L$ in the confining (high-$J$) phase, and very close to $0$ already in the deconfined phase.
Identifying the transition along the increasing-$\Gamma$ direction using $R(L)$ is difficult as the expectation value for both the Wilson loop and the horseshoe are exponentially small in $L$ and close to $0$, thus leading to large statistical errors in their ratio.
For practical purposes, one should instead use the \emph{dual} Wilson loop and horseshoes (defined as products of $\sigma^x$ in our model) to diagnose the transition along this direction.

\end{document}